\documentclass{aastex62}

\usepackage{amsmath}
\usepackage{units}

\begin{document}
\title{Stellar rotation and the extended main sequence turnoff in the open cluster NGC 5822}

\author{Weijia Sun} 
\affiliation{Kavli Institute for Astronomy \& Astrophysics and
  Department of Astronomy, Peking University, Yi He Yuan Lu 5, Hai
  Dian District, Beijing 100871, China}
  
\author{Richard de Grijs}
\affiliation{Department of Physics and Astronomy, Macquarie
  University, Balaclava Road, Sydney, NSW 2109, Australia}
\affiliation{International Space Science Institute--Beijing, 1
  Nanertiao, Hai Dian District, Beijing 100190, China}
  
\author{Licai Deng}
\affiliation{Key Laboratory for Optical Astronomy, National
  Astronomical Observatories, Chinese Academy of Sciences, 20A Datun
  Road, Chaoyang District, Beijing 100012, China}
\affiliation{School of Astronomy and Space Science, University of the
  Chinese Academy of Sciences, Huairou 101408, China}
\affiliation{Department of Astronomy, China West Normal University,
  Nanchong 637002, China}
  
\author{Michael D. Albrow}
\affiliation{School of Physical and Chemical Sciences, University of
  Canterbury, Private Bag 4800, Christchurch, New Zealand}

\begin{abstract}
The origin of extended main sequence turnoffs (eMSTOs) in
intermediate-age ($\unit[1-3]{Gyr}$) clusters is one of the most
intriguing questions in current star cluster research. Unlike the
split main sequences found in some globular clusters. which are caused
by bimodal populations in age and/or chemical abundances, eMSTOs are
believed to be owing to stellar rotation. We present a spectroscopic
survey of MSTO stars in a nearby, intermediate-age (\unit[0.9]{Gyr}),
low mass ($\sim\unit[1.7\times10^3]{M_\odot}$) Galactic open cluster,
NGC 5822. We derive a clean sample of member stars based on
\textit{Gaia} proper motions and parallaxes and confirm the existence
of an eMSTO. Using medium-resolution ($R\sim 4000$) Southern African
Large Telescope (SALT) spectra, we derive the rotational velocities of
24 member stars (representing 20\% completeness around the eMSTO
region) and find that the loci of the main sequence stars in the eMSTO
region show a clear correlation with the projected rotational
velocities in the sense that fast rotators are located on the red side
of the eMSTO and slow rotators are found on the blue side. By
comparison with a synthetic cluster model, we show that the stellar
rotational velocities and the eMSTO of NGC 5822 can be well reproduced
and we conclude that stellar rotation is the main cause of the eMSTO
in NGC 5822.
\end{abstract}


\section{Introduction \label{sec:intro}}

The extended main sequence turnoff (eMSTO) phenomenon---i.e., the
notion that the main sequence turnoff (MSTO) in the color--magnitude
diagram (CMD) is much wider than the prediction from single stellar
population modeling---first discovered in NGC 1846 by
\citet{2007MNRAS.379..151M}, is a common feature found in a large
fraction of young and intermediate-age ($\leqslant\unit[2]{Gyr}$)
massive Large and Small Magellanic Cloud clusters
\citep[e.g.][]{2008ApJ...681L..17M, 2009A&A...497..755M,
  2011ApJ...737....3G, 2014ApJ...784..157L, 2017MNRAS.467.3628C,
  2015MNRAS.450.3750M}.

In the past few years, our understanding of the eMSTO phenomenon in
these clusters has been enriched and enhanced significantly. Rather
than owing to intrinsic age spreads, stellar rotation is believed to
play an important role in shaping the morphology of the eMSTO
\citep[e.g.,][]{2014Natur.516..367L, 2015MNRAS.453.2070N,
  2009MNRAS.398L..11B}. This theory is further reinforced by multiple
lines of photometric and spectroscopic evidence. Using narrow-band
photometry, \citet{2017MNRAS.465.4795B} detected a large fraction
($\sim 30$--60\%) of Be stars in the MSTO regions of NGC 1850
($\sim\unit[80]{Myr}$) and NGC 1856 ($\sim\unit[280]{Myr}$), favoring
the interpretation that their split main sequences are caused by the
effects of fast rotators. Similar mechanisms were later confirmed in
NGC 1866 ($\sim\unit[200]{Myr}$) and NGC 1818 ($\sim\unit[40]{Myr}$)
through high-resolution spectroscopic surveys, suggesting that these
clusters host a blue main sequence composed of slow rotators and a red
one composed of fast rotators \citep{2017ApJ...846L...1D,
  2018AJ....156..116M}.

Aided by \textit{Gaia} Data Release 2 (DR2), we can derive `clean'
samples of open clusters in the Milky Way free of contamination by
field stars \citep{2018A&A...618A..93C}. The discovery of eMSTOs in
Galactic open clusters, similar to those observed in Magellanic Cloud
clusters, has opened up a new chapter in our comprehension of the
formation of eMSTOs and the evolution of open
clusters. \citet{2018ApJ...869..139C} found the existence of eMSTOs in
all 12 open clusters younger than $\sim\unit[1.5]{Gyr}$ they analyzed
(including NGC 5822), suggesting that eMSTOs are a common
feature of intermediate-age clusters in the Milky Way and that they
are regulated by the same mechanism as that operating in Magellanic
Cloud clusters. The understanding of open clusters, which used to be
considered prototypes of a single stellar population, encountered a
significant upheaval due to the discovery of eMSTOs in Galactic open
clusters. It has been reported that stars exhibit a wide range of
  rotation rates both in the field \citep{2010ApJ...722..605H} and in
  open clusters \citep{2006ApJ...648..580H}. A number of studies have
  explored the effect of stellar rotation on the CMD around the MSTO
  region \citep{2015ApJ...807...58B, 2015ApJ...807...25B,
    2015ApJ...807...24B}. However, the direct connection between the
  stellar rotation rates of MSTO stars and their loci in the CMD was
  only revealed recently. \citet{2018MNRAS.480.3739B} used Very Large
Telescope/FLAMES spectroscopy of 60 cluster members in NGC 2818, an
$\unit[800]{Myr}$ open cluster, to measure the stellar rotational
velocities and found that stars exhibiting high rotational velocities
are located on the red side of the eMSTO and those rotating slowly on
the blue side, in agreement with the prediction of the stellar
rotation scenario. \citet{2018ApJ...863L..33M} also reported that the
multiple sequences they found in the young cluster NGC 6705 correspond
to stellar populations with different rotation rates.

In this paper, we present a spectroscopic survey of MSTO stars in the
nearby ($\sim\unit[760]{pc}$) intermediate-age (\unit[0.9]{Gyr}) open
cluster NGC 5822. We find the presence of an eMSTO in this cluster and
verify that it is not an artifact caused by differential
extinction. The loci of the main sequence stars in the eMSTO region
show a clear correlation with the projected rotational velocities,
with fast rotators lying on the red side of the eMSTO and slow
rotators on the blue side. By comparison with a synthetic cluster
within the framework of stellar rotation, we argue that the observed
morphology of the eMSTO in the CMD can be properly explained by the
model and that stellar rotation is likely the main contributor to the
eMSTO morphology in NGC 5822.

This article is organized as follows. In Section \ref{sec:data} we
present the observations, data reduction procedures, and membership
determination. Section \ref{sec:eMSTO} reports our main results,
showing a strong correlation between the stellar rotation rates and
their loci in the CMD region covered by the eMSTO. A discussion and
our conclusions are summarized in Section \ref{sec:discussion}.

\section{Data Reduction and Analysis \label{sec:data}}
\subsection{Spectroscopic data}

We selected spectroscopic candidates in NGC 5822 using the photometric
survey in the $UBVI$ and $uvbyCa\mathrm{H}\beta$ systems undertaken by
\citet{2011AJ....142..127C}. These broad-band observations were
obtained with the Y4KCAM camera mounted on the Cerro Tololo
Inter-American Observatory (CTIO) \unit[1]{m} telescope and the
intermediate- and narrow-band imaging was carried out using the CTIO
\unit[0.9]{m} telescope. Through a cross-correlation with the UCAC3
database \citep{2000AJ....120.2131Z}, these authors derived 136
probable photometric members and 322 probable non-members.

We obtained spectroscopic observations with the Southern African Large
Telescope \citep[SALT;][]{2006SPIE.6267E..0ZB} equipped with the
Robert Stobie Spectrograph (RSS) using its multi-object spectroscopy
(MOS) capability over 9 nights from 2018 February 8/9 to 2018 August
14/15 under programs 2017-2-SCI-038 and 2018-1-SCI-006. Six masks were
designed to cover 88 stars (including repetitions) in NGC 5822
as part of program 2017-2-SCI-038, with three masks observed for the
second time the following semester (see Table \ref{tab:obs}). The
PG2300 grating was used with a \unit[1]{arcsec} wide short slit binned
$2\times 2$, offering a nominal spectral resolution of $\sim 4000$
with a per-pixel resolution of \unit[0.33]{\AA} at a central
wavelength of \unit[4884.4]{\AA}. Regular bias, argon arc lamp, and
quartz lamp flat field calibration frames were taken as part of normal
SALT operations. We used the PySALT package
\citep{2010SPIE.7737E..25C} to perform the primary reduction and
wavelength calibration. For all of our samples, we obtained spectra
with a signal-to-noise ratio (SNR) per pixel in excess of 200.

\begin{deluxetable*}{ccCCCCl}[b!]
\tablecaption{Observation Log of the SALT Runs \label{tab:obs}}
\tablewidth{0pt}
\tablehead{
\colhead{Mask Name} & \colhead{Programme} &
\colhead{$\alpha_\mathrm{J2000}$} &\colhead{$\delta_\mathrm{J2000}$} &\colhead{$N$\tablenotemark{a}} & \colhead{Exp. Time (s)} & \colhead{Date (UT)}  \\
\colhead{(1)} & \colhead{(2)} &
\colhead{(3)} & \colhead{(4)} & \colhead{(5)}& \colhead{(6)} &\colhead{(7)}
}
\startdata
NGC5822p2 &2017-2-SCI-038& 15^\mathrm{h}04^\mathrm{m}33.43^\mathrm{s}&-54\arcdeg26\arcmin33.16\arcsec & 13 & 600 & 2018 Feb 8 \\
&2018-1-SCI-006 & 15^\mathrm{h}04^\mathrm{m}33.43^\mathrm{s}&-54\arcdeg26\arcmin33.16\arcsec & 13 & 764 & 2018 Aug 3 \\
NGC5822p3 &2017-2-SCI-038& 15^\mathrm{h}04^\mathrm{m}19.43^\mathrm{s}&-54\arcdeg16\arcmin44.52\arcsec & 10 & 600 & 2018 Apr 26 \\
NGC5822p4 &2017-2-SCI-038& 15^\mathrm{h}03^\mathrm{m}11.44^\mathrm{s}&-54\arcdeg16\arcmin22.93\arcsec & 11 & 600 & 2018 Feb 8 \\
&2018-1-SCI-006& 15^\mathrm{h}03^\mathrm{m}11.44^\mathrm{s}&-54\arcdeg16\arcmin22.93\arcsec & 11 & 764 & 2018 Jul 30 \\
NGC5822p6 &2017-2-SCI-038& 15^\mathrm{h}03^\mathrm{m}09.95^\mathrm{s}&-54\arcdeg31\arcmin46.39\arcsec & 9 & 600 & 2018 Feb 11 \\
&2018-1-SCI-006& 15^\mathrm{h}03^\mathrm{m}09.95^\mathrm{s}&-54\arcdeg31\arcmin46.39\arcsec & 9 & 764 & 2018 Aug 14 \\
NGC5822p8 &2017-2-SCI-038& 15^\mathrm{h}05^\mathrm{m}10.13^\mathrm{s}&-54\arcdeg19\arcmin41.91\arcsec & 7 & 600 & 2018 Feb 26 \\
NGC5822p9 &2017-2-SCI-038& 15^\mathrm{h}04^\mathrm{m}36.35^\mathrm{s}&-54\arcdeg34\arcmin51.48\arcsec & 5 & 600 & 2018 Apr 30 \\
\enddata 
\tablenotetext{a}{Number of science slits in each field.}  
\end{deluxetable*}

\subsection{Membership determination}

\begin{figure}[ht!]
\gridline{\fig{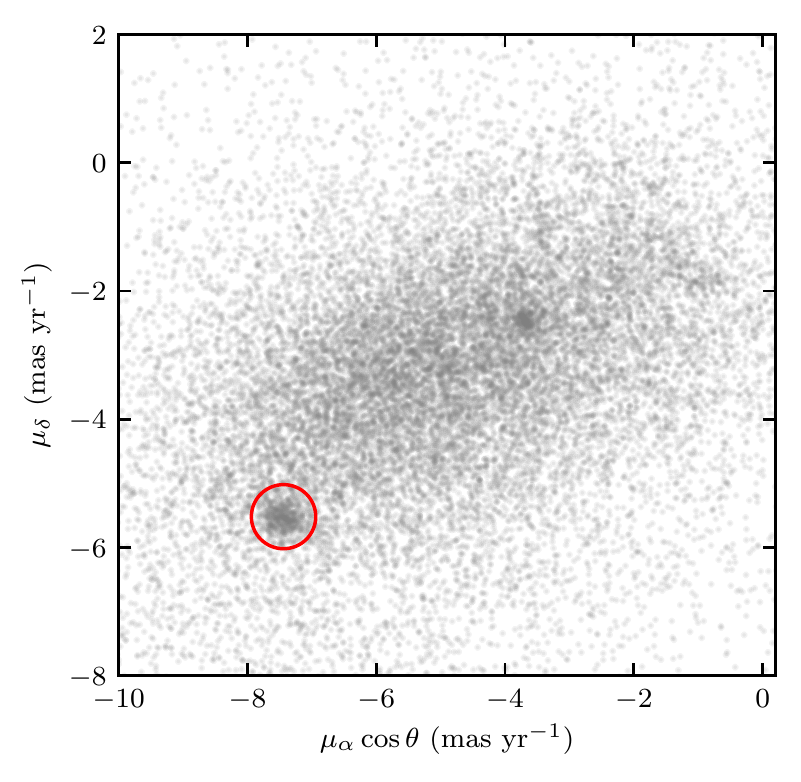}{0.5\textwidth}{}
          \fig{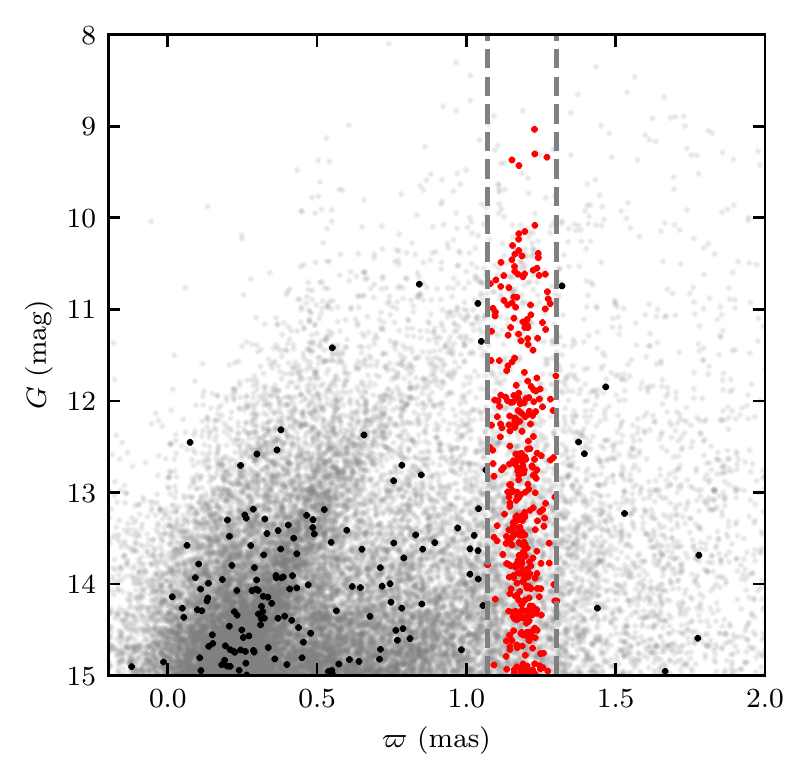}{0.5\textwidth}{}
}
\gridline{\fig{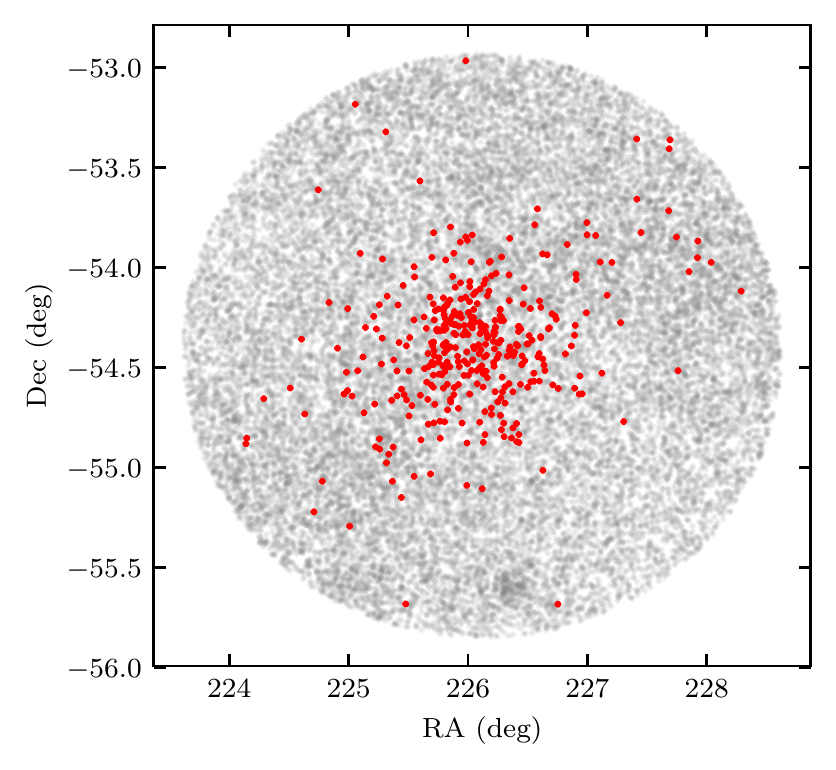}{0.5\textwidth}{}
          \fig{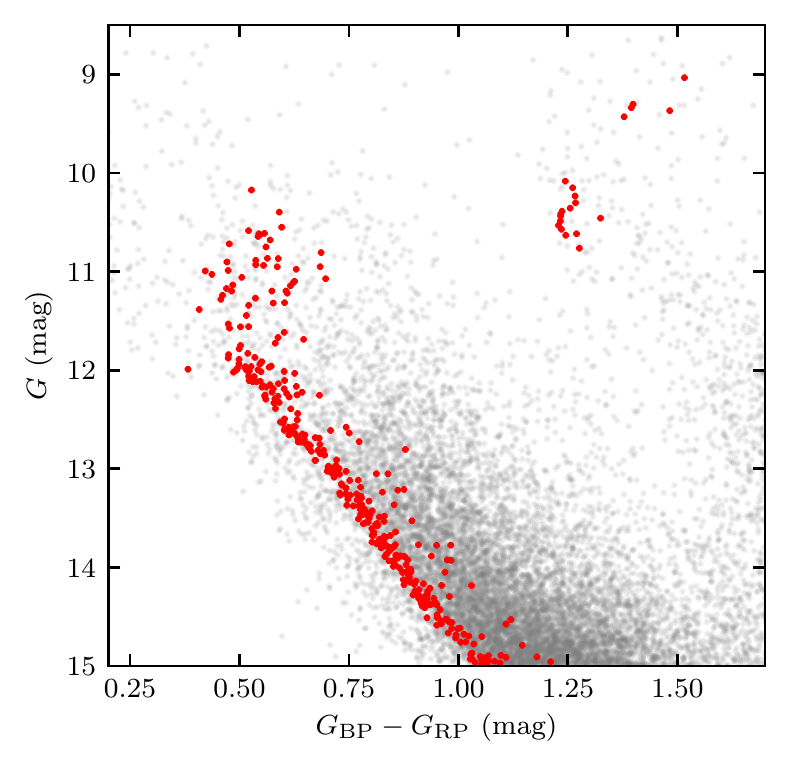}{0.5\textwidth}{}
}
\caption{{\bf (top left)} Vector-point diagram of the proper motions
  for stars brighter than $G = \unit[15]{mag}$ within 87.5\arcmin of
  the center of NGC 5822. The red circle shows the primary selection
  of cluster members. {\bf (top right)} $G$ band vs. parallaxes. The
  primary members selected from the proper motions are marked as solid
  dots and the parallax-selected members are marked as red dots. The
  vertical dashed lines represent the selection criteria applied to
  the parallaxes. {\bf (bottom left)} Spatial distribution of stars
  selected, with cluster members highlighted as red points. {\bf
    (bottom right)} CMD of all stars in the field (gray dots) and
  member stars of NGC 5822 (red solid dots). The eMSTO is visible
  around $G\sim\unit[11.5]{mag}$. \label{fig:selection}}
\end{figure}

\begin{figure}[ht!]
\plotone{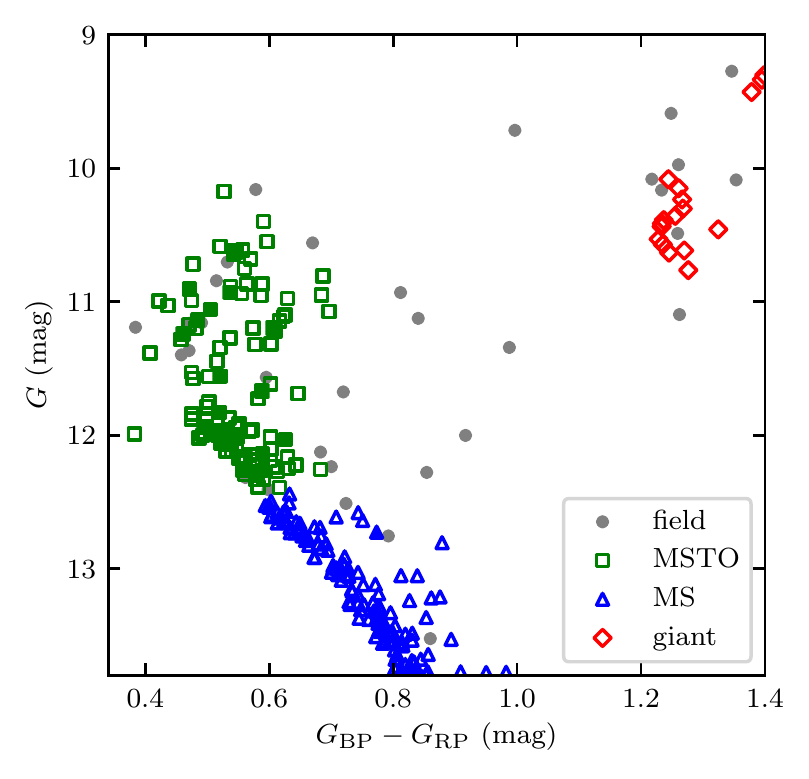}
\caption{CMD of NGC 5822 color-coded by the stellar
    classifications based on their loci. Member stars classified as
    MSTO, MS, and giant stars are marked as green squares, blue
    triangles, and red diamonds, respectively. Member stars with
    spectroscopic data are presented using solid markers and field
    stars with spectroscopic data are shown as gray
    circles. \label{fig:classification}}
\end{figure}

We exploited the \textit{Gaia} DR2 \citep{2016A&A...595A...1G,
  2018A&A...616A...1G} to analyze the stellar photometry, proper
motions, and parallaxes, and to perform membership determination in
the NGC 5822 field. First, we acquired the stellar catalog from the
\textit{Gaia} database within 2.5 times the cluster radius
\cite[35\arcmin;][]{2002A&A...389..871D}. In the vector-point diagram
(VPD) of stellar proper motions, NGC 5822 showed a clear concentration
centered at $(\mu_\alpha\cos\theta, \mu_\delta) \approx \unit[(-7.44,
  -5.52)]{mas\,yr^{-1}}$. The other overdensity located at
$(\mu_\alpha\cos\theta, \mu_\delta) \approx \unit[(-3.67,
  -2.52)]{mas\,yr^{-1}}$ corresponds to a nearby cluster, NGC
5823. Then, we derived the quantity $\mu_\mathrm{R} =
\sqrt{(\mu_\alpha\cos\theta-<\mu_\alpha\cos\theta>)^2+(\mu_\delta-<\mu_\delta>)^2}$
and applied a cut of $\mu_\mathrm{R} = \unit[0.4]{mas\,yr^{-1}}$ to
conduct our primary membership selection. Next, we placed a further
constraint on the parallaxes by estimating the mean parallax of the
proper-motion-selected stars
($\langle\varpi\rangle=\unit[1.18]{mas\,yr^{-1}}$) and adopted stars
with parallaxes within $\unit[0.115]{mas\,yr^{-1}}$ as cluster
members. Note that this approach is slightly different from that
  adopted by \citet{2018ApJ...869..139C}, in the sense that we adopted
  a straight cut in both $\mu_\mathrm{R}$ and $\varpi$ rather than
  applying different selection criteria for stars of different
  brightnesses. One reason for this approach is that NGC 5822 is
  sufficiently close that its member stars can be easily separated
  from field stars using parallaxes (see the top right-hand panel of
  Fig. \ref{fig:selection}). On the other hand, the limited number of
  stars in NGC 5822 makes it hard to reliably calculate the
  corresponding rms for each magnitude bin. Since we did not set out
  to compile a homogeneous database for multiple clusters, our
  approach is suitable for our analysis of this single cluster. We
present the spatial distribution as well as the CMD of the member
stars of NGC 5822, together with all stars in the field, in the bottom
panels of Fig. \ref{fig:selection}. We present the CMD of NGC
  5822 color-coded by the stellar classifications based on their loci
  in Fig. \ref{fig:classification}. Member stars classified as MSTO,
  MS, and giant stars are marked as green squares, blue triangles, and
  red diamonds, respectively. Member stars with spectroscopic data are
  presented using solid markers and field stars with spectroscopic
  data are shown as gray circles. Following decontamination of the
  field stars, 24 member stars (21 MSTO and 3 MS stars) were left in
  our observational sample; 13 member stars were observed a second
  time. We estimate that the total number of member MSTO stars in this
  cluster is $\sim$107, suggesting that the completeness of our
  observed sample is around 20\% in the eMSTO region.

\begin{figure}[ht!]
\plotone{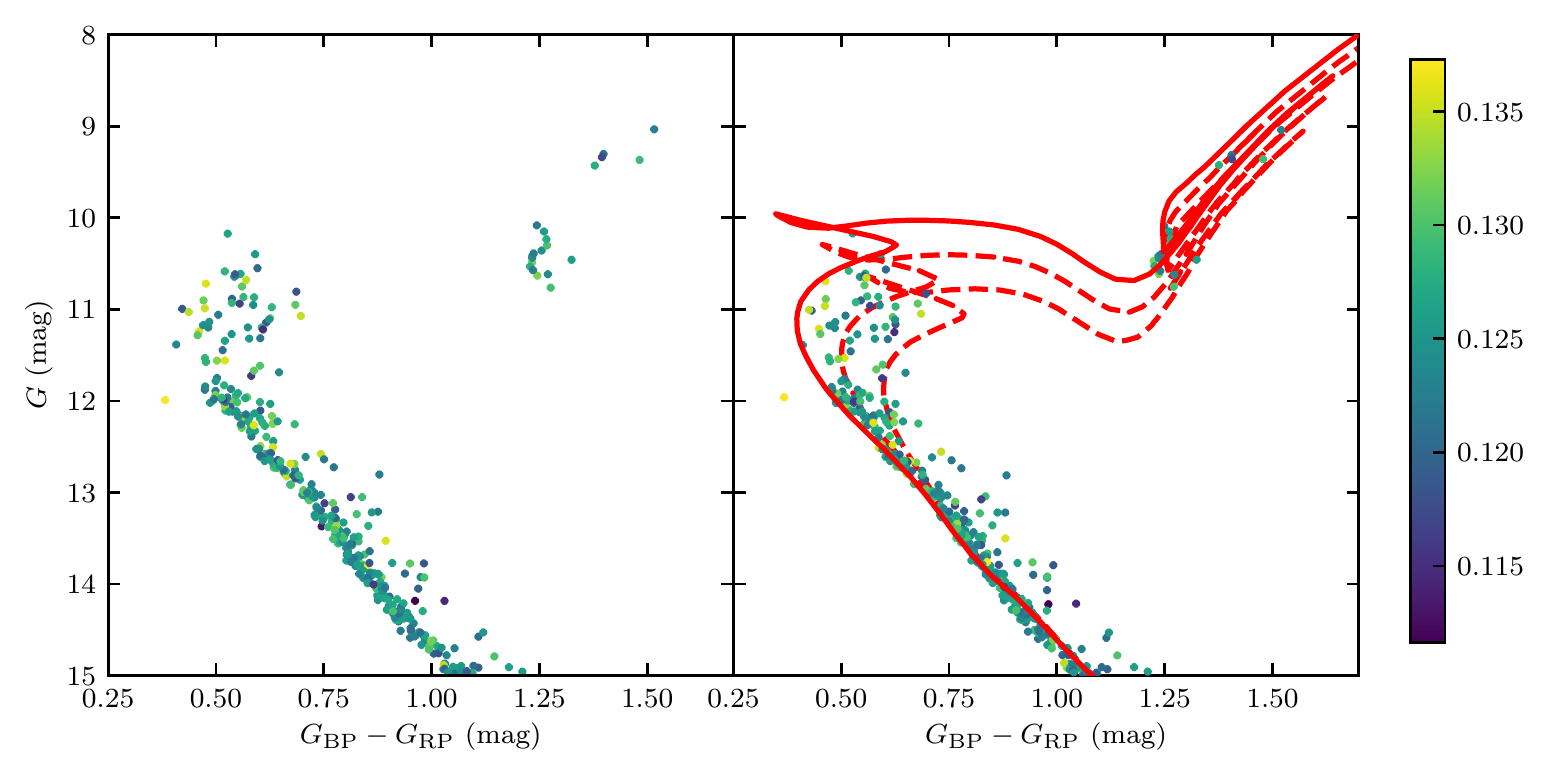}
\caption{Comparison of the CMD of selected cluster member stars before
  (left) and after (right) differential reddening
  correction. Different colors represent different extinctions
  $E(B-V)$. The best-fitting isochrone for the bulk stellar population
  is shown in the right-hand panel as a red solid line. The
  best-fitting isochrone (red solid line) has an age of
  \unit[0.9]{Gyr} with $Z = 0.017$ and a distance modulus of around
  9.4 ($\unit[\sim 760]{pc}$). Isochrones for ages of $\log( t \mbox{
    yr}^{-1})= 9.05$ and 9.15 are also overplotted (red dashed
  lines). \label{fig:cmd}}
\end{figure}

This cluster shows a clear eMSTO feature around
$G\sim\unit[11.5]{mag}$. To further demonstrate that this is not an
artifact owing to residual differential reddening, we estimated the
degree of the spatial variation of the reddening and found that its
influence is negligible compared with the extent of the eMSTO. Given
its close distance and low Galactic latitude, we found that we could
not use a 2D reddening map \citep[e.g.][]{2011ApJ...737..103S} to
estimate the differential reddening. Instead, we adopted the method of
\citet{2013ApJ...769...88N}, who assumed a two-component model for the
distribution of the dust, including the mean density of dust along the
plane $\rho_\mathrm{D}$ and a scale height $H_\mathrm{D}$. Therefore,
the prediction for the reddening in a given direction is given by
\begin{equation}
	E(B-V)=\rho_\mathrm{D}\int_0^d\exp(-r\sin(|b|)/H)\,\mathrm{d}r,
\end{equation}
where $b$ represents Galactic latitude and $d$ is the distance derived
from the corresponding \textit{Gaia} parallax. The distance was
derived from the parallax by implementing the formalism of
\citet{2016ApJ...832..137A}. $H=\unit[164]{pc}$ is the dust scale
height and $\rho_\mathrm{D}=\unit[0.427]{mag\,kpc^{-1}}$
\citep{2013ApJ...769...88N}. The average extinction for cluster
members is around $E(B-V)=\unit[0.126]{mag}$ with a standard deviation
of $\sigma_{E(B-V)} = \unit[0.004]{mag}$. Using the
\citet{1989ApJ...345..245C} and \citet{1994ApJ...422..158O} extinction
curve with $R_V = 3.1$, we corrected the reddening to the average
reddening value. In Fig. \ref{fig:cmd} we present the CMD of selected
cluster member stars before (left) and after (right) our differential
reddening correction. A visual inspection suggests that the morphology
of the CMD remains unchanged and the eMSTO still exists after having
applied this correction.

We used the Padova group's PARSEC 1.2S isochrones
\citep{2012MNRAS.427..127B} to perform our CMD fits based on visual
matching. \citet{2018ApJ...869..139C} derived an age of
$\sim\unit[1]{Gyr}$ and a solar-like metallicity ($Z_\odot =
0.0152$). Our best fit agrees with these results
(Fig. \ref{fig:cmd}). The best-fitting isochrone has an age of
\unit[0.9]{Gyr} for $Z = 0.017$ and a distance of $\unit[\sim
  760]{pc}$. The binary sequence is clearly visible in the CMD and
\citet{2018ApJ...869..139C} estimated the fraction of unresolved
binaries with $q>0.7$ at 0.131.
 
\subsection{Rotational velocities}

The projected rotational velocities were measured by fitting the
absorption line profiles of H$\beta$ and the Mg {\sc i} triplet. We
compiled a library of high-resolution synthetic stellar spectra with
effective temperatures, $T_\mathrm{eff}$, ranging from \unit[5000]{K}
to \unit[8000]{K} (in steps of \unit[100]{K}), surface gravities from
$\log g = 3.5$ to $\log g = 5.0$ (in steps of 0.1), and metallicities
from [Fe/H] = $-1.0$ to [Fe/H] = $1.0$ dex (in steps of 0.5 dex) from
the Pollux database \citep{2010A&A...516A..13P}. We applied the latest
ATLAS12 model atmospheres \citep{2005MSAIS...8..189K} where blanketed
model atmospheres handle line opacity in stellar atmospheres using the
Opacity Sampling technique. The models assume a plane parallel
geometry, hydrostatic and radiative equilibrium, as well as local
thermodynamic equilibrium. The microturbulent velocity was fixed to
$\unit[2]{km\,s^{-1}}$ for all models. Synthetic spectra were then
generated using the SYNSPEC tool \citep{1992A&A...262..501H}. Each
model spectrum was convolved with the rotational profile for a given
rotational velocity and implemented with an instrumental broadening as
well as a radial velocity shift. Given that the light enters through
off-axis slits (in the dispersion direction) in the MOS, the actual
resolution may vary from slit to slit and from mask to mask. Therefore
we adopted the full width at half maximum (FWHM) of the corresponding
arc lines as an indicator of the instrumental broadening effect. Then,
we used the Markov chain Monte Carlo
\citep[emcee;][]{2013PASP..125..306F} method to sample the
five-dimensional parameter space ($v\sin i$, $v_\mathrm{r}$,
$T_\mathrm{eff}$, $\log g$, [Fe/H]) to employ a $\chi^2$ minimization.
For each of the 3000 runs of the MCMC procedures, $\chi^2$ values and
their associated probabilities $e^{-\chi^2/2}$ were
stored. Probability distributions were then generated by projecting
the sum of the probabilities onto the dimension considered. A Gaussian
fit to the distribution provides its width $\sigma$, which we adopt as
the uncertainty.

\begin{figure}[ht!]
\plotone{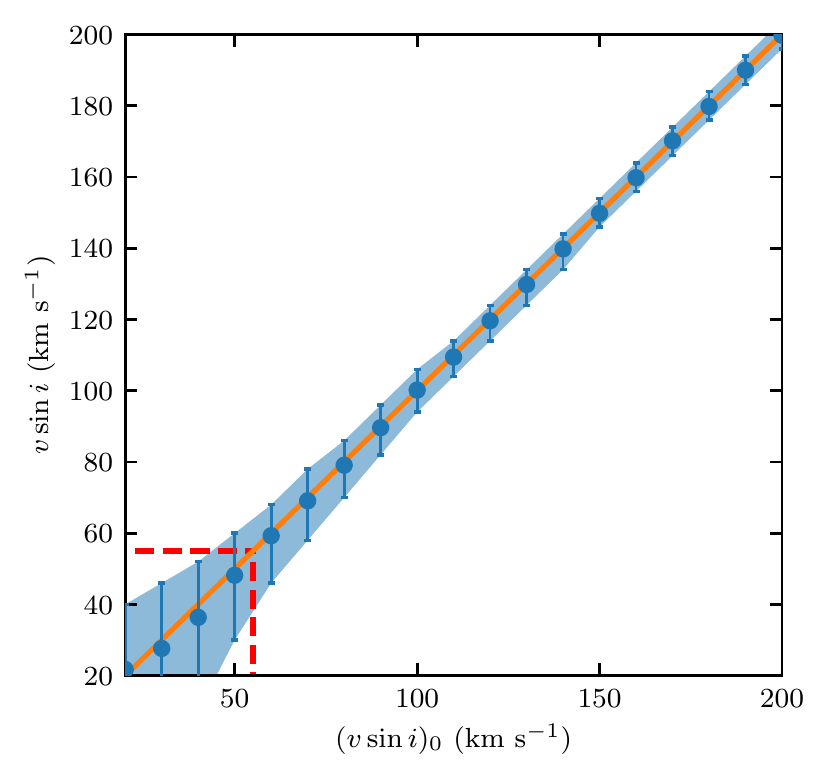}
\caption{Rotational velocities of the mock data (horizontal axis)
  vs. rotational velocities derived from profile fitting. The blue
  shadowed region corresponds to $\unit[1]{\sigma}$ and the one-to-one
  relation is indicated by an orange solid line. The red dashed lines
  represent the lower limit of reliable measurements of the rotational
  velocity ($\unit[55]{km\,s^{-1}}$) where the uncertainty is around
  half of the actual value. \label{fig:error}}
\end{figure}

To estimate the influence of instrumental broadening on the
determination of the rotational velocities, we generated a set of mock
spectra by sampling the projected rotational velocities $v\sin i$ from
$\unit[20]{km\,s^{-1}}$ to $\unit[200]{km\,s^{-1}}$, assuming a
uniform $\mathrm{SNR} = 200$ and a reasonable uncertainty for the
instrumental broadening ($\sigma_\mathrm{FWHM} = \unit[0.1]{\AA}$),
and we measured the best-fitting parameters from those mock
spectra. We repeated this procedure 100 times and estimated the median
values and the 68th percentiles of the velocity distribution. In
Fig. \ref{fig:error} we present a comparison of the rotational
velocities of the mock data with those derived through profile
fitting. The blue shadowed region corresponds to $\unit[1]{\sigma}$
and the one-to-one relation is indicated by an orange solid
line. Given the intermediate spectral resolution, it is hard to
differentiate the effect of rotational from instrumental broadening
for slow rotators. Therefore, we defined the detection limit of the
rotational velocity to be the velocity where its uncertainty is around
half of the actual value and, for slow rotators with $v\sin i
\leqslant \unit[55]{km\,s^{-1}}$, the uncertainties of the
measurements are comparable to their actual values, while the
uncertainty is less than 5\% and 3\% for the mock spectra with $v\sin
i\geqslant \unit[100]{km\,s^{-1}}$ and $v\sin i\geqslant
\unit[150]{km\,s^{-1}}$, respectively.

\section{Extended MSTOs and stellar rotation \label{sec:eMSTO}}

The eMSTO of NGC 5822, if interpreted as an age difference, is around
$\unit[300-350]{Myr}$. \citet{2018ApJ...869..139C} estimated the ages
of the stars around the eMSTO region by linearly interpolating a grid
of isochrones and calculated the FWHM of the cluster's age
distribution, which gives a spread of $\unit[270\pm52]{Myr}$. They
also showed that the FWHM of the NGC 5822 eMSTO follows the
correlation between the width of the eMSTO and cluster age applicable
within the framework of stellar rotation.

\begin{figure}[ht!]
\gridline{\fig{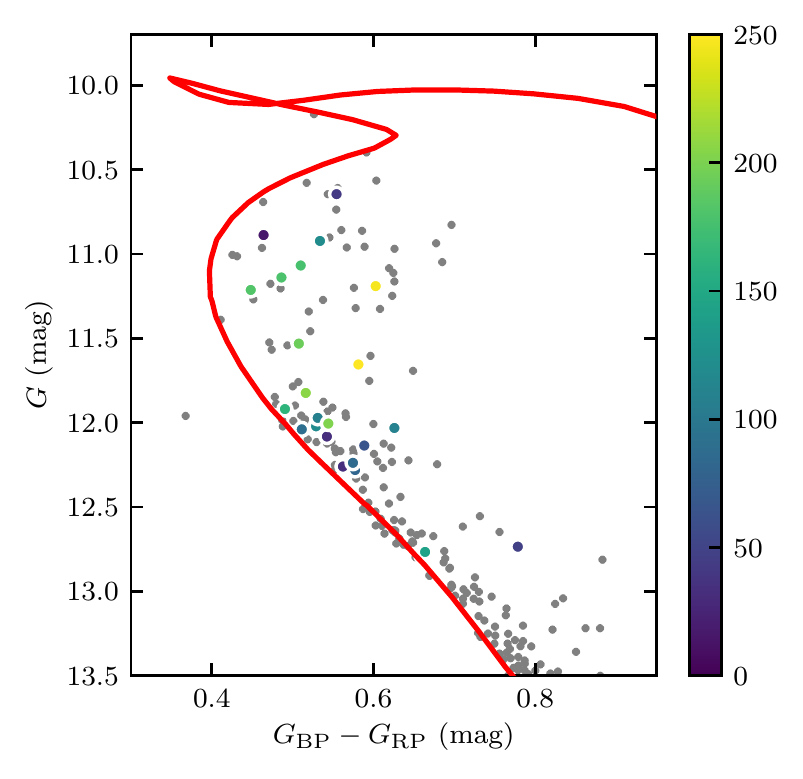}{0.4\textwidth}{}
          \fig{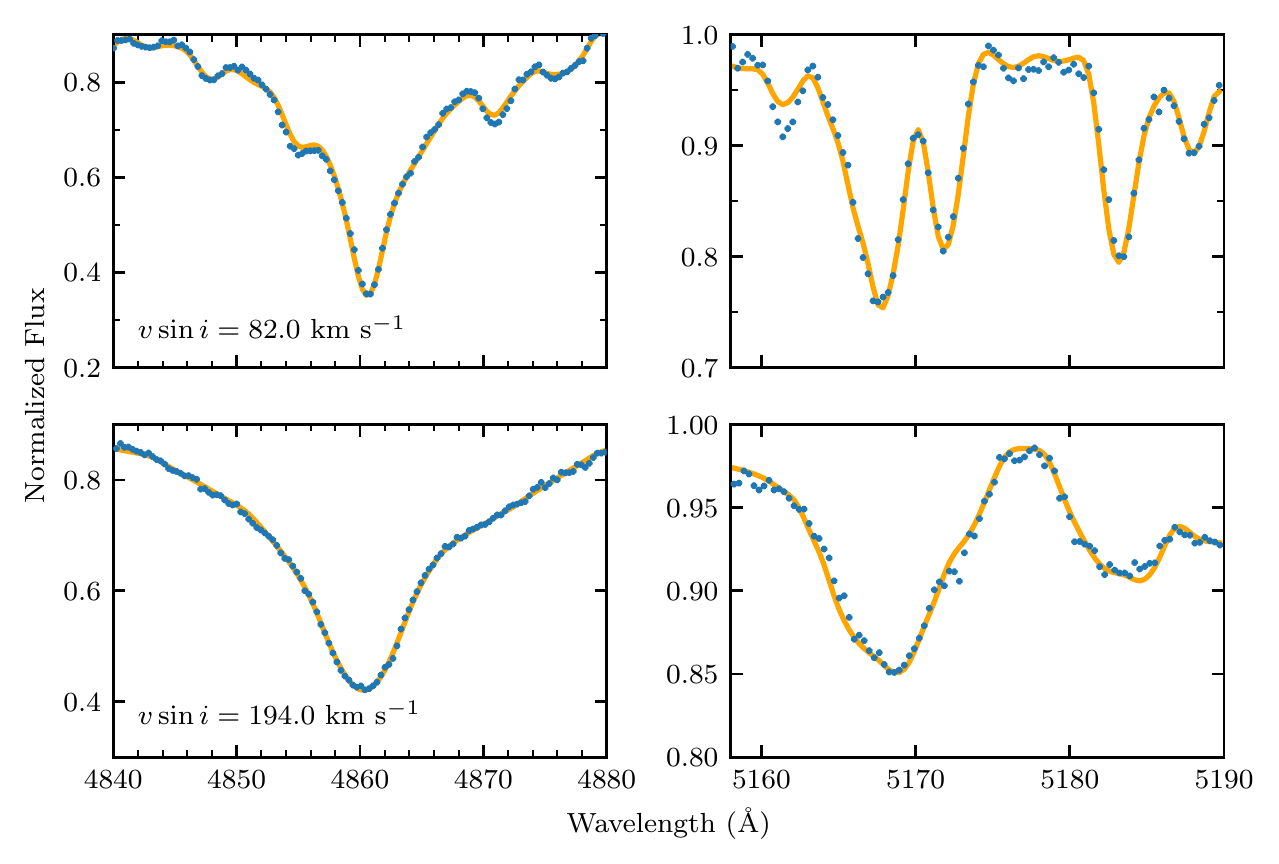}{0.6\textwidth}{}
}
\caption{{\bf (left)} CMD of NGC 5822 with the member stars
  color-coded by their rotational velocities. The best-fitting
  isochrone is shown as the red curve. A clear trend between stellar
  rotation and their loci in the CMD region is seen, in the sense that
  the rapid rotators (yellow) tend to lie on the red side of the eMSTO
  while the slow rotators (blue) are usually found on the blue
  side. {\bf (right)} {\bf Two sample spectra of a slow rotator (top)
    and a fast rotator (bottom). H$\beta$ and Mg {\sc i} triplets of
    the same object are shown in the left- and right-hand columns,
    respectively. For each spectrum, the best-fitting models are
    presented as orange curves.} \label{fig:rot}}
\end{figure}

In Fig. \ref{fig:rot}, we present the CMD of NGC 5822, with the member
stars color-coded by their rotational velocities. We found that their
loci in the CMD region covered by the eMSTO strongly depend on stellar
rotation, in the sense that rapid rotators tend to lie on the red side
of the eMSTO while slow rotators are usually found on the blue
side. Similar results have also been discovered in young and
intermediate-age clusters in the Magellanic Clouds
\citep{2017ApJ...846L...1D, 2018MNRAS.480.1689K}, as well as in
Galactic open clusters \citep{2018ApJ...863L..33M,
  2018MNRAS.480.3739B}. The stellar structural parameters as well as
the inferred projected rotational velocities are listed in Table
\ref{tab:res}.

\begin{deluxetable*}{lCCCRRRRR}[b!]
\tablecaption{Properties of member stars with rotational velocity measurements. \label{tab:res}}
\tablewidth{0pt}
\tablehead{
\colhead{\text{Gaia} ID}&\colhead{$G$ (mag)}&\colhead{$G_\mathrm{bp}$ (mag)}&\colhead{$G_\mathrm{rp}$ (mag)} &\colhead{$T$ (K)\tablenotemark{a}} &\colhead{$\log g$\tablenotemark{a}} & \colhead{Fe/H\tablenotemark{a}} & \colhead{$v\sin i$ ($\unit{km\,s^{-1}}$)\tablenotemark{b}} & \colhead{$v_\mathrm{rv}$ ($\unit{km\,s^{-1}}$)\tablenotemark{c}}  \\
\colhead{(1)} & \colhead{(2)} &
\colhead{(3)} & \colhead{(4)} & \colhead{(5)}& \colhead{(6)} &\colhead{(7)}&\colhead{(8)}&\colhead{(9)}
}
\startdata
5887641469783582208 & 12.28 & 12.50 & 11.92& 6900\pm 100 & 3.5\pm 0.1 & 0.0\pm 0.5 & 82.0\pm12.4 & -17.8\pm5.8 \\ 
5887666444964220416 & 11.21 & 11.38 & 10.93& 7400\pm 100 & 3.9\pm 0.1 & 0.0\pm 0.5 & 182.3\pm6.1 & -30.2\pm7.0 \\ 
5887666444964220416\tablenotemark{*} & 11.21&11.38 & 10.93& 7200\pm100 & 3.6\pm0.1 & 0.0\pm0.5 & 176.9\pm6.1 & -25.2\pm5.1 \\ 
5887669198096568960 & 12.03 & 12.27 & 11.64& 6800\pm 100 & 4.2\pm 0.1 & -0.5\pm 0.5 & 110.8\pm8.1 & -42.0\pm6.1 \\ 
5887668648340678784 & 10.89 & 11.07 & 10.61& 7200\pm 100 & 4.4\pm 0.1 & 0.0\pm 0.5 & 16.0\pm38.8 & -30.3\pm6.6 \\ 
5887642466216028032 & 11.82 & 12.02 & 11.50& 7100\pm 100 & 4.1\pm 0.1 & -0.5\pm 0.5 & 206.2\pm5.8 & -24.1\pm6.7 \\ 
5887642466216028032\tablenotemark{*} & 11.82&12.02 & 11.50& 7000\pm100 & 4.2\pm0.1 & -0.5\pm0.5 & 211.3\pm5.6 & -23.1\pm14.1 \\ 
5887668575267987840 & 11.92 & 12.10 & 11.61& 6900\pm 100 & 3.7\pm 0.1 & -0.5\pm 0.5 & 163.0\pm6.1 & -26.8\pm13.6 \\ 
5887698644394381952 & 10.65 & 10.87 & 10.31& 7200\pm 100 & 3.6\pm 0.1 & 0.0\pm 0.5 & 42.7\pm24.7 & -40.1\pm6.9 \\ 
5887698644394381952\tablenotemark{*} & 10.65&10.87 & 10.31& 7200\pm100 & 3.5\pm0.1 & 0.0\pm0.5 & 53.9\pm20.3 & -33.4\pm4.4 \\ 
5887719054073235584 & 11.07 & 11.27 & 10.76& 7400\pm 100 & 4.0\pm 0.1 & 0.0\pm 0.5 & 177.6\pm6.1 & -31.6\pm5.2 \\ 
5887719054073235584\tablenotemark{*} & 11.07&11.27 & 10.76& 7000\pm100 & 4.0\pm0.1 & -0.5\pm0.5 & 181.7\pm6.1 & -32.9\pm6.5 \\ 
5887644935768013056 & 11.53 & 11.72 & 11.21& 7400\pm 100 & 4.1\pm 0.1 & 0.0\pm 0.5 & 194.0\pm6.0 & -42.3\pm5.2 \\ 
5887644935768013056\tablenotemark{*} & 11.53&11.72 & 11.21& 7400\pm100 & 4.0\pm0.1 & 0.0\pm0.5 & 194.2\pm6.0 & -46.4\pm9.5 \\ 
5887671255327574144 & 12.08 & 12.29 & 11.75& 7000\pm 100 & 3.5\pm 0.1 & 0.0\pm 0.5 & 33.4\pm29.0 & -21.5\pm6.0 \\ 
5887695414549752704 & 10.92 & 11.13 & 10.60& 7300\pm 100 & 3.8\pm 0.1 & 0.0\pm 0.5 & 120.9\pm7.3 & -36.6\pm6.6 \\ 
5887695414549752704\tablenotemark{*} & 10.92&11.13 & 10.60& 7200\pm100 & 3.7\pm0.1 & 0.0\pm0.5 & 124.9\pm7.0 & -26.0\pm6.8 \\ 
5887718740485998208 & 11.19 & 11.43 & 10.83& 7000\pm 100 & 3.6\pm 0.1 & 0.0\pm 0.5 & 246.9\pm2.4 & -27.7\pm8.6 \\ 
5887718740485998208\tablenotemark{*} & 11.19&11.43 & 10.83& 7100\pm100 & 4.3\pm0.1 & -0.5\pm0.5 & 239.0\pm3.5 & -33.2\pm7.7 \\ 
5887668197310871168 & 12.04 & 12.23 & 11.72& 7100\pm 100 & 3.7\pm 0.1 & 0.0\pm 0.5 & 89.0\pm11.0 & 15.9\pm7.1 \\ 
5887642981611923200 & 11.66 & 11.88 & 11.30& 7000\pm 100 & 4.2\pm 0.1 & -0.5\pm 0.5 & 249.1\pm2.1 & -28.2\pm6.8 \\ 
5887642981611923200\tablenotemark{*} & 11.66&11.88 & 11.30& 6900\pm100 & 3.5\pm0.1 & -0.5\pm0.5 & 249.6\pm2.0 & -26.5\pm9.7 \\ 
5887722313898780672 & 12.02 & 12.22 & 11.69& 7400\pm 100 & 4.3\pm 0.1 & 0.0\pm 0.5 & 120.0\pm7.3 & -31.9\pm5.2 \\ 
5887722313898780672\tablenotemark{*} & 12.02&12.22 & 11.69& 7300\pm100 & 4.1\pm0.1 & 0.0\pm0.5 & 143.0\pm6.3 & -26.9\pm8.1 \\ 
5887671397119565312 & 12.74 & 13.04 & 12.26& 6300\pm 100 & 4.0\pm 0.1 & -0.5\pm 0.5 & 48.2\pm22.4 & -28.6\pm8.4 \\ 
5887698296441666688 & 11.97 & 12.17 & 11.64& 7400\pm 100 & 3.9\pm 0.1 & 0.5\pm 0.5 & 108.2\pm8.3 & -31.9\pm5.0 \\ 
5887698296441666688\tablenotemark{*} & 11.97&12.17 & 11.64& 7400\pm100 & 3.7\pm0.1 & 0.5\pm0.5 & 83.7\pm12.0 & -29.3\pm5.5 \\ 
5887697987204027264 & 11.14 & 11.33 & 10.85& 7400\pm 100 & 3.8\pm 0.1 & 0.0\pm 0.5 & 178.9\pm6.1 & -31.6\pm8.9 \\ 
5887697987204027264\tablenotemark{*} & 11.14&11.33 & 10.85& 7400\pm100 & 3.9\pm0.1 & 0.0\pm0.5 & 175.1\pm6.1 & -38.0\pm6.1 \\ 
5887671534558688640 & 12.79 & 13.04 & 12.38& 6500\pm 100 & 3.7\pm 0.1 & -0.5\pm 0.5 & 138.4\pm6.4 & -21.8\pm7.8 \\ 
5887670224535430272 & 12.26 & 12.47 & 11.91& 6600\pm 100 & 3.5\pm 0.1 & -0.5\pm 0.5 & 34.1\pm28.7 & -58.6\pm8.3 \\ 
5887671431479631360 & 12.77 & 13.02 & 12.36& 6600\pm 100 & 3.7\pm 0.1 & -0.5\pm 0.5 & 144.0\pm6.3 & -30.3\pm7.0 \\ 
5887665895208406272 & 12.01 & 12.21 & 11.66& 7000\pm 100 & 3.9\pm 0.1 & -0.5\pm 0.5 & 201.6\pm5.9 & -32.7\pm8.1 \\ 
5887665895208406272\tablenotemark{*} & 12.01&12.21 & 11.66& 7000\pm100 & 4.4\pm0.1 & -0.5\pm0.5 & 202.4\pm5.9 & -42.0\pm6.7 \\ 
5887667544475841152 & 12.14 & 12.36 & 11.77& 6800\pm 100 & 4.0\pm 0.1 & -0.5\pm 0.5 & 65.2\pm16.6 & -51.4\pm4.4 \\ 
5887642805464232704 & 12.24 & 12.45 & 11.87& 7200\pm 100 & 4.3\pm 0.1 & 0.5\pm 0.5 & 86.0\pm11.6 & -14.4\pm9.7 \\ 
5887642805464232704\tablenotemark{*} & 12.24&12.45 & 11.87& 7000\pm100 & 4.3\pm0.1 & 0.5\pm0.5 & 123.5\pm7.1 & -0.0\pm13.6 \\ 
\enddata 
\tablenotetext{a}{Uncertainty adopted from the step size of the model grid.}
\tablenotetext{b}{Uncertainty estimated from the mock test.}
\tablenotetext{c}{Uncertainty given by the MCMC procedure.}
\tablenotetext{*}{Duplicate observation in Programme 2018-1-SCI-006}  
\tablecomments{(1) \textit{Gaia} DR2 ID; (2,3,4) Extinction-corrected \textit{Gaia} bands (5) Effective temperature; (6) Surface gravity; (7) Metallicity; (8) Projected rotational velocity; (9) Radial velocity.}
\end{deluxetable*}

\begin{figure}[ht!]
\gridline{\fig{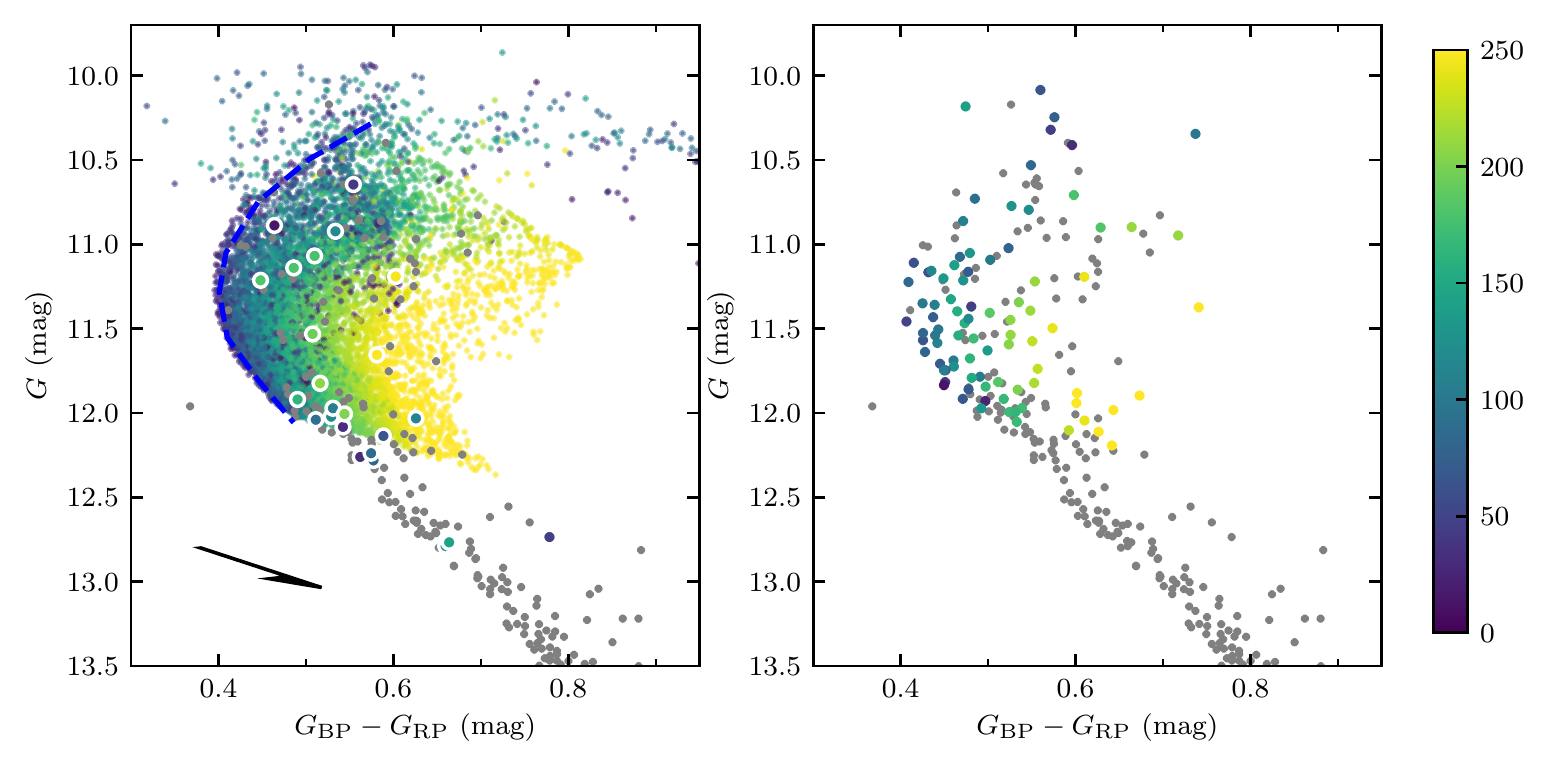}{0.66\textwidth}{}
          \fig{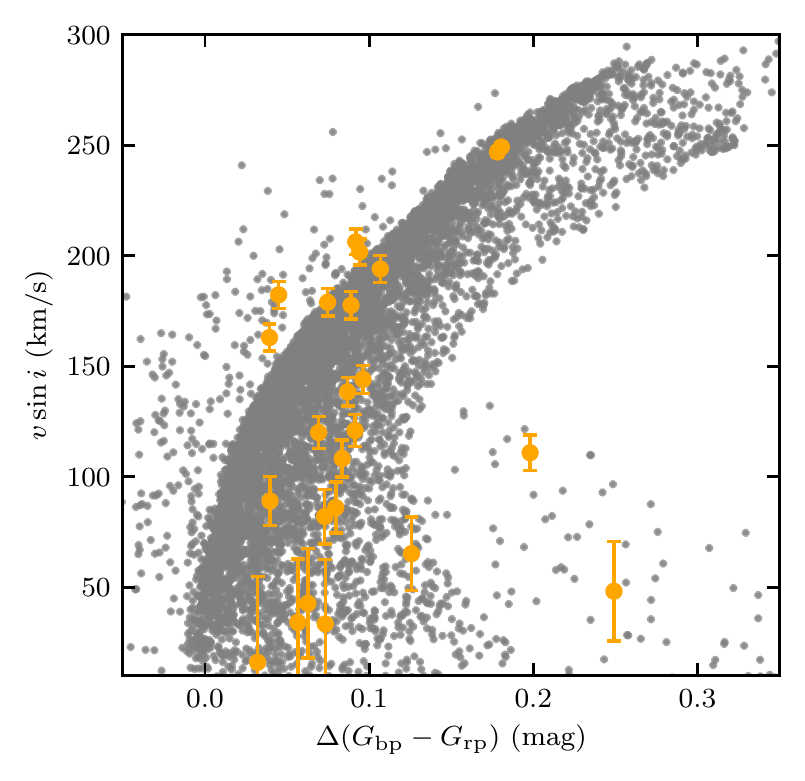}{0.34\textwidth}{}
}
\caption{{\bf (left)} CMD of the observed cluster (gray dots), as well
  as the synthetic cluster, with colors representing the projected
  rotational velocities. Larger points with white borders reflect
  measurements of member stars. The adopted ridgeline is shown as a
  blue dashed line. The black arrow represents indicates how stellar
  rotation affects the locus of a star in the CMD. {\bf (middle)} {\bf
    Realistic synthetic cluster with a number of stars comparable to
    that in the observed CMD.} {\bf (right)} Pseudo-color vs.
  rotational velocity. Stars with projected rotational velocity
  measurements are marked as orange dots and the synthetic cluster
  stars are marked as gray dots.\label{fig:model}}
\end{figure}

We also compared the observed cluster data with a synthetic cluster
data set that included the effects of stellar rotation. The synthetic
cluster data waere derived from the SYCLIST models
\citep{2013A&A...553A..24G, 2014A&A...566A..21G}, assuming a
metallicity of $Z=0.014$, an age of $\log( t \mbox{ yr}^{-1}) = 8.95$,
and a binary fraction of 0.131, with a rotational distribution derived
from \citet{2010ApJ...722..605H} and a random rotation axis
distribution. The model also accounts for the limb-darkening effect
\citep{2000A&A...359..289C} as well as for the gravity darkening law
of \citet{2011A&A...533A..43E}. In the left-hand panel of
Fig. \ref{fig:model}, the synthetic cluster is superposed onto the CMD
of NGC 5822 and the eMSTO feature is well-reproduced and consistent
with coeval stellar populations with different rotation rates. The
projected rotational velocity of the synthetic cluster follows a
similar trend as the real member stars, which become redder as the
stellar rotation rates increase. In the middle panel we present a
  realistic synthetic cluster with a number of stars comparable to
  that in the observed CMD.

To provide a better comparison with the simulation, we introduced the
pseudo-color $\Delta{(G_\mathrm{bp}-G_\mathrm{rp})}$ as the normalized
color difference with respect to the blue ridgeline in the direction
determining how stellar rotation may change the locus of a star in the
CMD (black arrow) to represent the deviation in color which may be
caused by stellar rotation. We adopted the blue edge of the synthetic
cluster, which represents the population of non-rotating stars, as the
fiducial ridgeline. In the right-hand panel of Fig. \ref{fig:model},
the $\Delta{(G_\mathrm{bp}-G_\mathrm{rp})}$ vs. $v\sin{i}$ diagram for
all stars with projected rotational velocity measurements is shown,
and the gray dots represent the same distribution for the synthetic
cluster. We found that most of our targets follow the trend predicted
by the stellar rotational model, where the pseudo-color is close to
zero for slow rotators and it increases significantly as the
rotational velocity increases. Two outliners in the right-hand panel
of Fig. \ref{fig:model} (\textit{Gaia} ID: 5887669198096568960 and
5887671397119565312), which have relatively large pseudo-colors
compared with their rotational velocities, may result from
contamination by binary stars. Since their locations in the CMD
coincide with the equal-mass binary sequence, they are likely
unresolved binaries, particularly the star with \textit{Gaia} ID
5887671397119565312, whose low mass ($\unit[1.2-1.3]{M_\odot}$) is
close to the minimum mass for large stellar rotation. With such a low
mass, stars brake efficiently early on the MS and evolve back to the
non-rotating tracks \citep{2018arXiv181205544G}. Therefore, stellar
rotation is unlikely the cause of such a large shift in color and
binary stars might be a plausible explanation of these two outliners.

\section{Discussion and conclusions\label{sec:discussion}}

NGC 5822 is an intermediate-age (\unit[0.9]{Gyr}) Galactic open
cluster exhibiting an eMSTO. Through membership determination based on
\textit{Gaia} proper motions and parallaxes, we investigated the CMDs
of NGC 5822 and confirmed that the eMSTO is unlikely an artifact
caused by differential extinction. By exploiting SALT/RSS data, we
derived the projected rotational velocities of 24 member stars and
found that stellar rotation is strongly correlated with the stellar
loci in the CMD in the MSTO region. The red side of the eMSTO is
occupied by fast rotators while the blue side is mainly composed of
slow rotators. By comparison with a synthetic cluster, we have shown
that the eMSTO of NGC 5822 can be properly reproduced and the
rotational velocities of the eMSTO stars follow the same pattern as
that predicted by the stellar rotation model.

\begin{figure}[ht!]
\plotone{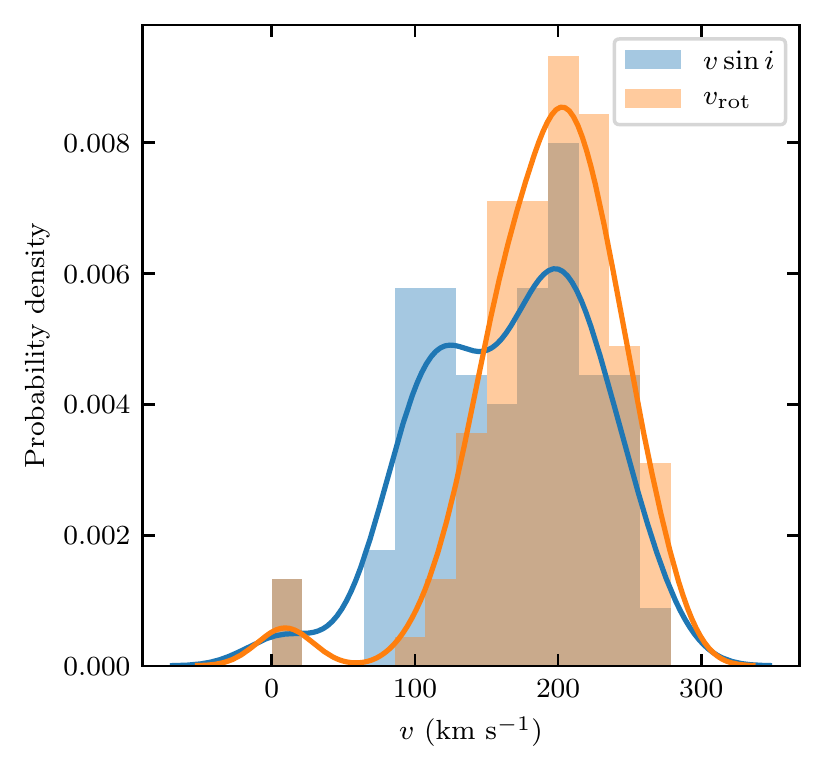}
\caption{Distributions of projected rotational velocities $v\sin i$
  (blue) and true rotational velocities $v_\mathrm{rot}$ (orange) for
  member stars in the MSTO region of NGC 5822. Both velocities were
  estimated by taking the average velocities of the nearest 50 stars
  in the synthetic CMD. \label{fig:prob}}
\end{figure}

Combined with NGC 6705 \citep[250 Myr;][]{2018ApJ...863L..33M} and NGC
2818 \citep[800 Myr;][]{2018MNRAS.480.3739B}, we have confirmed the
correlation between stellar rotation and stellar positions in the
eMSTO/split MS in three Galactic open
clusters. \citet{2018ApJ...869..139C} found that Galactic open
clusters also follow the trend between cluster age and the extent of
eMSTO seen in Magellanic Cloud clusters \citep{2013ApJ...776..112Y},
suggesting that they should be regulated by a similar mechanism. Split
MSs are believed to be composed of two stellar populations
characterized by different rotation rates: the blue MS is composed of
slow rotators and the red MS is composed of rapid rotators
\citep{2016MNRAS.458.4368M}. Spectroscopic surveys of the young
clusters NGC 1818 \citep{2018AJ....156..116M} and NGC 6705
\citep{2018ApJ...863L..33M} confirmed the existence of slowly and
rapidly rotating populations and found that these two subgroups are
well separated in projected rotational velocity, with a difference in
mean $v\sin i$ greater than $\unit[100]{km\,s^{-1}}$. Meanwhile, in
the intermediate-age clusters NGC 2818 \citep{2018MNRAS.480.3739B} and
NGC 5822, such a result is barely seen, which may due to small-number
statistics and selection effects. Therefore, we estimated the
rotational velocities for all MSTO stars in NGC 5822 based on our
synthetic cluster to check the distribution of the stellar rotation
rates. On the basis of previous analyses, we argue that the synthetic
cluster can properly reproduce the observed results and that it can be
taken as a reasonable approximation to the real cluster. Thus, for
each member star in the NGC 5822 MSTO region, we inferred the
rotational velocities by taking the average velocities of the nearest
50 stars in the synthetic CMD.  

The distributions of projected rotational velocity $v\sin i$ and
$v_\mathrm{rot}$ are presented in Fig. \ref{fig:prob}. We found that
the projected rotational velocities show a dip around
$\unit[150]{km\,s^{-1}}$, similar to the results for NGC 1818 and NGC
6705. However we suggest that this is an artifact caused by projection
effects. In Fig. \ref{fig:prob}, the `slow' rotators have a peak at
$\unit[100]{km\,s^{-1}}$ and they have a dearth of stars with $v\sin i
\sim \unit[50]{km\,s^{-1}}$, which is different from the results for
the young clusters where the slowly rotating populations have lower
mean velocities and do not show a gap in slowly rotating stars. The
distribution of the true rotational velocities is also shown in
Fig. \ref{fig:prob} and the fact that it shows a single peak around
$\unit[200]{km\,s^{-1}}$ further confirms that the equatorial
velocities in NGC 5822 should follow a unimodal distribution. On the
other hand, projection effects are unlikely to explain the large
difference in projected rotational velocities found in young clusters
and the true rotation rates of MSTO stars in young clusters should in
all probability show a bimodal distribution given the fact that the
split MSs can be separated into distinct sequences in the CMD. Since
the typical masses of the split MSs in young clusters
($\geqslant\unit[2.5]{M_\odot}$) and eMSTOs in intermediate-age
clusters ($\unit[1.4]{M_\odot}$--$\unit[2]{M_\odot}$) are different,
if a split MS and eMSTO are present in the evolutionary sequence of
star clusters, these two distributions of stellar rotation may coexist
in the same cluster. This may hint that the stellar rotation
distribution in clusters follow a similar pattern as the field
population in the sense that stars more massive than
$\unit[2.5]{M_\odot}$ show a bimodal equatorial velocity distribution
while less massive stars have a unimodal rotation distribution
\citep{2012A&A...537A.120Z}. However, spin alignment in clusters may
play an overlooked role. \citet{2017NatAs...1E..64C} found evidence of
spin alignment among the red giant stars in the two old open clusters
NGC 6791 and NGC 6819. \citet{2018NatAs.tmp..156L} inferred the
$v_\mathrm{rot}$ and the inclination angles $i$ of MSTO stars in NGC
6705 from Monte Carlo simulations where $v_\mathrm{rot}$ has a linear
distribution and $i$ a Gaussian distribution. They argued that cluster
members have highly aligned spin axes, which implies a link between
stellar rotation and rotational kinetic energy in the progenitor
molecular cloud.

We also estimated the stellar masses following similar procedures as
\citet{2018ApJ...862..133S}. In essence, we generated a synthetic
cluster of 10,000 stars with the initial masses generated through
Monte Carlo sampling of a Kroupa stellar initial mass function
\citep[IMF;][]{2001MNRAS.322..231K}. Then, we calculated the ratio of
the number of member stars with $G$ magnitudes from \unit[12.5]{mag}
to \unit[13.5]{mag} to that of the synthetic cluster for the same
magnitude range. We multiplied the integrated mass obtained for the
synthetic cluster by this ratio to estimate the total stellar mass in
the cluster, $\unit[1.7\pm0.3\times10^3]{M_\odot}$. We confirmed that
changing the magnitude range will not affect the estimation of the
total mass significantly. \citet{2018MNRAS.480.3739B} reported a mass
of $\unit[2800]{M_\odot}$ for NGC 2818. These results suggest that
eMSTOs are not exclusive to massive clusters
($\unit[10^4-10^5]{M_\odot}$).

One possible source that could also give rise to a broadened MSTO is
stellar variability. \citet{2016ApJ...832L..14S} argued that the
instability strip intersects with the MSTO region of a cluster with an
age of $\unit[\sim 1-3]{Gyr}$ and can make a significant contribution
to the observed eMSTOs. Follow-up observations of NGC 1846 revealed
the presence of a group of (mainly $\delta$ Scuti) variable stars
around the eMSTO region. However, the number fraction of variable
stars was not sufficient to produce the observed width of the eMSTO
\citep{2018AJ....155..183S}. Certain types of variable or binary stars
(e.g., EA-type eclipsing binaries) exhibit large changes in radial
velocity over time and can be detected through multi-epoch
observations. However, we did not find such candidates in our sample
because of the limited spectral resolution and the small number of
observations. Follow-up photometric and spectroscopic observations are
required to further investigate ithe role of variability in shaping
the morphology of the eMSTO region.

\acknowledgments R. d. G. and L. D. acknowledge research support from
the National Natural Science Foundation of China through grants
11633005, 11473037, and U1631102. R. d. G. is grateful for support
from the National Key Research and Development Program of China
through grant 2017YFA0402702 from the Chinese Ministry of Science and
Technology (MOST). L. D. also acknowledges support from MOST through
grant 2013CB834900.

\vspace{5mm}
\facilities{SALT(RSS)}

\software{PySALT \citep{2010SPIE.7737E..25C}, PARSEC \citep[1.2S;][]{2012MNRAS.427..127B}, Astropy \citep{2013A&A...558A..33A}, Matplotlib \citep{2007CSE.....9...90H}, SYNSPEC \citep{1992A&A...262..501H}, emcee \citep{2013PASP..125..306F}}


\end{document}